%% file: 0_final_paper.tex
\documentclass[conference]{IEEEtran}
\IEEEoverridecommandlockouts
\usepackage{cite}
\usepackage{amsmath,amssymb,amsfonts}
\usepackage{algorithm}
\usepackage{algorithmic}
\usepackage{graphicx}
\usepackage{bbm}
\usepackage{textcomp}
\usepackage{xcolor}
\usepackage{latexsym}
\usepackage{gensymb}
\usepackage{siunitx}
\usepackage{tikz}
\usepackage{etoolbox}
\usepackage{enumitem}
\usepackage{esvect}
\usepackage{booktabs}
\usepackage{mathtools}
\usetikzlibrary{chains,arrows,arrows.meta,calc,positioning}
\usepackage{multirow}
\usepackage{comment}
\usepackage[normalem]{ulem}
\usepackage{upquote}
\usepackage{caption}
\usepackage{subcaption}
\def\BibTeX{{\rm B\kern-.05em{\sc i\kern-.025em b}\kern-.08em
    T\kern-.1667em\lower.7ex\hbox{E}\kern-.125emX}}

\def\Exp{\mathbb{E}\,}

\newcommand{\bs}{\boldsymbol}
\newcommand{\mc}{\mathcal}
\newcommand{\wh}{\widehat}

\newcommand{\herm}{^{\text{\sf H}}}

\newcommand{\CNorm}{\mc{N}_{\mathbb{C}}}

\newcommand{\bkt}[1]{\langle {#1} \rangle}

\usepackage{float}
\begin{document}

\IEEEoverridecommandlockouts
\addtolength{\topmargin}{0.06in}
\addtolength{\textheight}{-0.06in}

\title{Learning-Based Signal Recovery in Nonlinear Systems with Spectrally Separated Interference
}

\author{
\IEEEauthorblockN{Jayadev Joy and Sundeep Rangan}
\IEEEauthorblockA{Department of Electrical and Computer Engineering,\\
New York University, Brooklyn, NY, USA\\
}}

\maketitle

\begin{abstract}
Upper Mid-Band (FR3, 7–24 GHz) receivers for 6G must operate over wide bandwidths in dense spectral environments, making them particularly vulnerable to strong adjacent-band interference and front-end nonlinearities. While conventional linear receivers can suppress spectrally separated interferers under ideal hardware assumptions, receiver saturation and finite-resolution quantization cause nonlinear spectral leakage that severely degrades performance in practical wideband radios. We study the recovery of a desired signal from nonlinear receiver observations corrupted by a high-power out-of-band interferer. The receiver front-end is modeled as a smooth, memoryless nonlinearity followed by additive noise and optional quantization. To mitigate these nonlinear and quantization-induced distortions, we propose a learned multi-layer Vector Approximate Message Passing (LMLVAMP) algorithm that incorporates spectral priors with neural network based denoising. Simulation results demonstrate significant performance gains over conventional methods, particularly in high-interference regimes representative of FR3 coexistence scenarios.
\end{abstract}


\input{1_intro}

\input{2_system_model}

\input{3_signal_recovery}

\input{4_performance}

\input{5_conclusion}

\bibliographystyle{IEEEtran}
\bibliography{ref}

\end{document}

%% file: 1_intro.tex
\section{Introduction}
Modern wireless communication systems increasingly operate in spectrally dense environments, where coexistence with strong adjacent-band signals is inevitable~\cite{intf}. This is particularly critical in emerging upper mid-band (FR3, 7–24 GHz) deployments that must support diverse services such as satellite and terrestrial cellular networks~\cite{kang2024cellular,testolina2024sharing}. Conventional receivers rely on linear front-ends and frequency-selective filtering~\cite{oppen}. While such filtering can ideally remove disjoint-band interference, nonlinear hardware effects such as amplifier saturation and analog-to-digital converter (ADC) clipping~\cite{razavi} cause spectral leakage, which are common in cost and energy efficient FR3 hardware. A major challenge arises when a strong out-of-band (OOB) interferer drives the receiver into a nonlinear regime. Even if the interferer lies outside the desired band, nonlinear distortion introduces in-band leakage, making classical filtering ineffective.

\subsection{Related Work}
\subsubsection{Classical Nonlinearity Modeling}  
Traditional mitigation uses polynomial modeling and inverse compensation. Volterra and memory polynomial models have been applied to Radio Frequency (RF) front-ends~\cite{volt}, while digital pre-distortion (DPD) is standard at transmitters~\cite{dpd} and has been extended to receivers. These methods demand accurate models of hardware nonlinearity and memory, which are difficult to obtain in practice, and adaptive estimation of high-order models is computationally prohibitive.

\subsubsection{Classical Interference Suppression}  
Strong OOB interferers can saturate Low Noise Amplifiers (LNAs), mixers, and ADCs. Analog filtering and automatic gain control (AGC) circuits help, but are insufficient in wideband or reconfigurable radios~\cite{reconrf}. Self-interference cancellation works in full-duplex systems, but requires careful hardware design and only cancels predictable interference. Digital methods such as pulse blanking and spectral notching fail when nonlinear saturation causes in-band leakage. Linearization via Bussgang’s theorem~\cite{Bussgang} reduces nonlinear effects to distortion noise, ignoring signal-dependent artifacts.

\subsubsection{Message Passing Algorithms}  
Approximate Message Passing (AMP)~\cite{amp} and Generalized AMP (GAMP)~\cite{gamp} enable scalable Bayesian inference but diverge under structured sensing. Vector AMP (VAMP)~\cite{vamp,vamp_glm} improves robustness with right-orthogonally invariant matrices, but requires accurate priors and models. Classical denoisers in these methods are fixed and struggle with nonlinearities such as saturation. Thus, while rigorous under structured priors, they are limited in handling practical hardware distortions.

\subsubsection{Data-Driven Approaches}  
Deep unfolding combines iterative structures with trainable modules. Learned VAMP~\cite{amp_dn}, Learned denoising-based AMP~\cite{ldamp}, and Orthogonal AMP-Net~\cite{hehe} have shown strong results in linear and low-resolution tasks. However, they are not designed for nonlinearities from front-end saturation or OOB interference, and often generalize poorly beyond training conditions.

\subsection{Motivation for Our Approach}
We address the recovery of signals corrupted by strong OOB interference, nonlinear saturation, and coarse quantization, impairments common in modern FR3 receiver front-ends. We propose a learned version of the multi-layer VAMP (ML-VAMP)~\cite{fletcher2018inference, pandit2020inference, manoel2017multi} framework, integrating model-based inference with data-driven adaptability. The nonlinear model accounts for receiver noise, smooth saturation, and optional quantization, while the estimator leverages spectral priors and trains its denoisers end-to-end to handle nonlinear effects. Unlike fixed model-based methods, the approach does not require explicit distortion knowledge, and unlike black-box networks, it retains interpretability via iterative inference.

Simulations show that the proposed estimator outperforms linear baselines, particularly in saturation-dominated regimes, representative of upper mid-band coexistence scenarios. These results highlight the importance of jointly modeling nonlinearity, noise, and quantization in the design of robust and cost-efficient FR3 receivers for future 6G systems.

%% file: 2_system_model.tex
\section{Problem Formulation}

\subsection{Spectral Signal Model}
We adopt the signal model of~\cite{dutta2023capacity}, a discrete-time approximation for systems with signals in distinct frequency bands. Let \( N \) be the number of time-domain samples in an observation period, and \( \bs{r} \in \mathbb{C}^N \) the received sample vector. Its frequency-domain decomposition is modeled as:
\begin{equation}
    \bs{r} = \sum_{\ell=0}^{L-1} \bs{V}\herm \bs{x}_\ell,
\end{equation}
where \( \bs{V}\herm \in \mathbb{C}^{N \times N} \) is the unitary inverse discrete Fourier transform (IDFT) matrix, \( L \) the number of signal sources, and \( \bs{x}_\ell \in \mathbb{C}^N \) the frequency-domain signal from source \( \ell \).

Each received signal is a complex Gaussian random vector with independent components:
\begin{equation} \label{eq:varx}
    x_\ell[i] \sim
    \begin{cases}
        \CNorm(\mu_\ell[i], S_\ell), & i \in B_\ell \\
        0, & \text{otherwise,}
    \end{cases}
\end{equation}
where \( B_\ell \subseteq \{0,\ldots,N-1\} \) denotes the interval of frequency bins of source \( \ell \), and \( \mu_\ell[i] \) and \( S_\ell \) are the prior mean and variance. The sources occupy disjoint bands, so that:
\begin{equation} \label{eq:Bdisjoint}
    B_\ell \cap B_k = \emptyset \quad \text{for } \ell \neq k.
\end{equation}
There are two common cases for the parameters \( \mu_\ell[i] \) and \( S_\ell \):
\begin{itemize}
    \item \emph{Unknown signal}: \( S_\ell > 0 \), \( \mu_\ell[i] = 0 \), i.e., a zero-mean complex Gaussian with variance \( S_\ell \).
    \item \emph{Known signal}: \( S_\ell = 0 \), \( \mu_\ell[i] \neq 0 \), representing known interference (e.g., decoded interferers).
\end{itemize}

\begin{figure}[t]
    \centering
    \input{figures/spec_model.tex}
    \caption{Spectral structure for \(L = 2\), where the desired user \(\bs{x}_0\) and interferer \(\bs{x}_1\) occupy disjoint bands \(B_0\) and \(B_1\). Bars show \(\mathrm{Var}[x[k]]\),  where \(\bs{x}\) is the DFT of \(\bs{r}\).}
    \label{fig:spectral}
\end{figure}
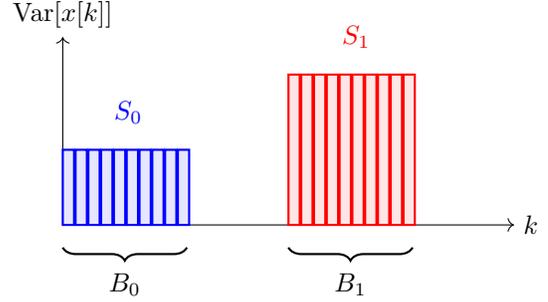

In simulations we set \( L=2 \): \( \ell=0 \) is the desired user and \( \ell=1 \) the interferer. Fig.~\ref{fig:spectral} shows a typical two-signal spectral configuration. The desired signal is always unknown, while the interferer may be known or unknown.

\subsection{General Nonlinear Model}
We model the receiver front-end using a general nonlinear measurement model~\cite{lozano2023spectral}:
\begin{equation} \label{eq:nonlin}
    \bs{y} = \Phi(\bs{r}, \bs{w}),
\end{equation}
where \( \bs{y} \) is the receiver output, \( \Phi(\cdot) \) a (possibly nonlinear) front-end function, and \( \bs{w} \) a noise vector (e.g., from thermal effects in the circuitry).

Assuming a \emph{memoryless} receiver, both \( \bs{y} \) and \( \bs{w} \) are \( N \)-dimensional, and each sample of~\eqref{eq:nonlin} is given by:
\begin{equation} \label{eq:nonlin_scal}
    y[i] = \phi(r[i], w[i]),
\end{equation}
where \( \phi(\cdot) \) is a scalar nonlinear function applied independently to each time sample. This captures the combined effect of components such as low-noise amplifiers, mixers, and quantizers in RF chains.

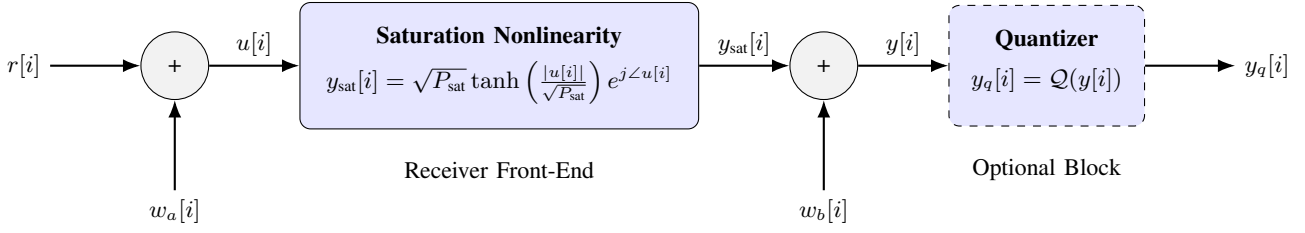
\begin{figure*}[t]
    \vspace{1mm}
    \centering
    \input{figures/sys_model.tex}
    \caption{Block diagram of a nonlinear receiver chain incorporating saturation, receiver noise, and an optional quantizer.}
    \label{fig:sys_model}
\end{figure*}

\subsection{Soft Saturation Model}
As a specific instance of~\eqref{eq:nonlin_scal}, we consider a soft-threshold model~\cite{lozano2023spectral}. The noise vector is given by:
\begin{equation} \label{eq:wsoft}
    w[i] = (w_a[i], w_b[i]), \;\; 
    w_j[i] \sim \CNorm(0, \sigma_j^2), \;\; j \in \{a,b\},
\end{equation}
with independent complex Gaussian components. The nonlinear measurement, illustrated in Fig.~\ref{fig:sys_model}, is given by:
\begin{equation} \label{eq:phi_sat}
    \phi(r, w) = f\!\left(\frac{|r+w_a|}{\sqrt{P_{\text{sat}}}}\right)(r+w_a) + w_b,
\end{equation}
where the index \( i \) is omitted for notational simplicity. Here, \( P_{\text{sat}} \) is the saturation power threshold, and $f(.)$ is defined as:
\begin{equation} \label{eq:f_x}
    f(x) = \frac{\tanh(x)}{x}.
\end{equation}
In practical systems with finite-resolution ADCs, the output is quantized:
\begin{equation} \label{eq:quant_phi}
    y_q[i] = \mathcal{Q}(y[i]), \quad 
    \phi_q(r[i], w[i]) = \mathcal{Q}\big(\phi(r[i], w[i])\big),
\end{equation}
where \( \mathcal{Q}(\cdot) \) represents a scalar uniform quantizer.

This memoryless nonlinearity models RF front-ends by compressing amplitude while preserving phase. While the proposed model is general, it is particularly relevant to FR3 receivers, where large instantaneous bandwidths and relaxed analog filtering lead to frequent exposure to high-power adjacent-band signals. The receiver is assumed to know \( S_0, S_1, B_0, B_1, P_{\text{sat}} \) (estimated via pilots or spectrum sensing), but not \( \phi(\cdot) \). The task is to estimate the desired signal \( \bs{x}_0 \) from \( \bs{y} \) or its quantized version \( \bs{y}_q \).

%% file: figures/spec_model.tex
\begin{tikzpicture}
\usetikzlibrary{decorations.pathreplacing}

\draw[->] (0,0) -- (6,0) node[right] {\(k\)};
\draw[->] (0,0) -- (0,2.5) node[above] {\(\mathrm{Var}[x[k]]\)};

\def\barwidth{0.15}
\def\gap{0.02}

\foreach \i/\h in {
    0/1.0, 1/1.0, 2/1.0, 3/1.0, 4/1.0,
    5/1.0, 6/1.0, 7/1.0, 8/1.0, 9/1.0
} {
    \pgfmathsetmacro{\x}{\i*(\barwidth+\gap)}
    \fill[blue!10] (\x,0) rectangle (\x+\barwidth,\h);
    \draw[blue, thick] (\x,0) rectangle (\x+\barwidth,\h);
}

\foreach \i/\h in {
    0/2.0, 1/2.0, 2/2.0, 3/2.0, 4/2.0,
    5/2.0, 6/2.0, 7/2.0, 8/2.0, 9/2.0
} {
    \pgfmathsetmacro{\x}{3.0 + \i*(\barwidth+\gap)}
    \fill[red!10] (\x,0) rectangle (\x+\barwidth,\h);
    \draw[red, thick] (\x,0) rectangle (\x+\barwidth,\h);
}

\node[blue] at (0.87,1.5) {\(S_0\)};
\node[red] at (3.9,2.5) {\(S_1\)};

\draw[decorate,decoration={brace,mirror,amplitude=5pt},thick] (0,-0.3) -- (1.65,-0.3)
node[midway, below=6pt] {\(B_0\)};
\draw[decorate,decoration={brace,mirror,amplitude=5pt},thick] (3.0,-0.3) -- (4.65,-0.3)
node[midway, below=6pt] {\(B_1\)};

\end{tikzpicture}

%% file: figures/sys_model.tex
\begin{tikzpicture}[node distance=1.6cm and 2.2cm, >=Latex]

\tikzset{
  block/.style = {draw, rectangle, rounded corners, minimum height=1.2cm, minimum width=2.6cm, align=center, font=\small, fill=blue!10},
  sum/.style = {draw, circle, minimum size=0.9cm, inner sep=0pt, font=\small, fill=gray!10},
  arrow/.style = {thick, ->},
  label/.style = {font=\small},
  explain/.style = {font=\small, align=center}}

  \node[font=\small, align=center] (r) at (0,0) {$r[i]$};

  \node[sum, right=1.2cm of r] (sumw0) {+};
  \node[below=1.2cm of sumw0, font=\small, align=center] (w0) {$w_a[i]$};
  \draw[arrow] (r) -- (sumw0) node[midway, above, label] {};
  \draw[arrow] (w0.north) -- (sumw0.south) node[midway, right, label] {};

  \node[block, minimum width=5.3cm, right=1.2cm of sumw0] (nonlin) {\\[2pt]
    \textbf{Saturation Nonlinearity}\\[5pt]
    $y_{\text{sat}}[i] = \sqrt{P_{\text{sat}}} \tanh\left(\frac{|u[i]|}{\sqrt{P_{\text{sat}}}}\right) e^{j \angle u[i]}$\\[2pt]};
  \node[explain, below=0.3cm of nonlin] {Receiver Front-End};
  \draw[arrow] (sumw0) -- (nonlin) node[midway, above, label] {$u[i]$};

  \node[sum, right=1.2cm of nonlin] (sumw1) {+};
  \node[below=1.2cm of sumw1, font=\small, align=center] (w1) {$w_b[i]$};
  \draw[arrow] (nonlin) -- (sumw1) node[midway, above, label] {$y_{\text{sat}}[i]$};
  \draw[arrow] (w1.north) -- (sumw1.south) node[midway, right,  label] {};

  \node[block, dashed, right=1.2cm of sumw1] (quant) {\\[2pt]
    \textbf{Quantizer}\\[5pt]
    $y_q[i] = \mathcal{Q}(y[i])$\\[2pt]};
  \node[explain, below=0.3cm of quant] {Optional Block};
  \draw[arrow] (sumw1) -- (quant) node[midway, above,  label] {$y[i]$};

  \node[right=1.2cm of quant, font=\small, align=center] (output) {$y_q[i]$};
  \draw[arrow] (quant) -- (output) node[midway, above,  label] {};

\end{tikzpicture}

%% file: 3_signal_recovery.tex
\section{Learned ML-VAMP for Nonlinear Filtering}

\subsection{Bayesian ML-VAMP Algorithm}
As introduced, the proposed algorithm is a ``learned'' version of ML-VAMP~\cite{fletcher2018inference, pandit2020inference, manoel2017multi}. Before describing it, we briefly review ML-VAMP for nonlinear interference. The measurements can be written as:
\begin{equation} \label{eq:yxmulti}
    \bs{y} = \Phi(\bs{r}, \bs{w}), \quad \bs{r} = \bs{V}\herm \bs{x},
\end{equation}
where \(\bs{x}\) is the frequency-domain vector and \(\bs{r}\) its time-domain representation. The goal is to estimate \(\bs{x}\) and \(\bs{r}\) from \(\bs{y}\). In ML-VAMP terminology, the system~\eqref{eq:yxmulti} is a two-layer network: the first layer maps \(\bs{x} \to \bs{r}\), and the second maps \(\bs{r} \to \bs{y}\). The algorithm iteratively updates estimates \(\wh{\bs{x}}^{(t)}\) and \(\wh{\bs{r}}^{(t)}\) for \(t = 0,1,\dots\) as:
\begin{subequations} \label{eq:mlvamp}
\begin{align}
    \wh{\bs{x}}^{(t)} &= G_0(\bs{z}_0^{(t)}, \gamma_0), \quad
    \alpha_0^{(t)} = \langle G_0'(\bs{z}_0^{(t)}, \gamma_0) \rangle \label{eq:mlvamp_g0}, \\
    \bs{z}_1^{(t)} &= \frac{1}{1 - \alpha_0^{(t)}} \bs{V}\herm \left( \wh{\bs{x}}^{(t)} - \alpha_0^{(t)} \bs{z}_0^{(t)} \right) \label{eq:mlvamp_z1}, \\
    \gamma^{(t)}_1 &= \gamma_0^{(t)}(1/\alpha_0^{(t)} - 1) \label{eq:mlvamp_gam1}, \\
    \wh{\bs{r}}^{(t)} &= G_1(\bs{z}_1^{(t)}, \gamma_1^{(t)}), \quad
    \alpha_1^{(t)} = \langle G_1'(\bs{z}_1^{(t)}, \gamma_1^{(t)}) \rangle \label{eq:mlvamp_g1},\\
    \bs{z}_0^{(t+1)} &= \frac{1}{1 - \alpha_1^{(t)}} \bs{V} \left( \wh{\bs{r}}^{(t)} - \alpha_1^{(t)} \bs{z}_1^{(t)} \right),\\
    \gamma^{(t+1)}_0 &= \gamma_1^{(t)}(1/\alpha_1^{(t)} - 1) \label{eq:mlvamp_gam0},
\end{align}
\end{subequations}
initialized with:
\begin{equation}
    \bs{z}_0^{(0)} = \bs{0}, 
    \quad \gamma^{(0)}_0 = 0.
\end{equation}

The updates in~\eqref{eq:mlvamp} use two functions, \(G_0(\cdot)\) and \(G_1(\cdot)\), called \emph{denoisers}. For OOB interference rejection, \(G_0(\cdot)\) is the \emph{spectral denoiser}, acting on the frequency-domain signal, and \(G_1(\cdot)\) is the \emph{nonlinear denoiser}, acting in the time domain where the nonlinearity occurs. In~\eqref{eq:mlvamp_g0} and~\eqref{eq:mlvamp_g1}, \(\langle \bs{v} \rangle\) denotes the empirical mean of \(\bs{v}\), defined as:
\begin{equation}
    \langle \bs{v} \rangle = \frac{1}{n} \sum_{i=0}^{n-1} v[i].
\end{equation}
Accordingly, the scalar terms \( \alpha_0^{(t)} \) and \( \alpha_1^{(t)} \) represent the average divergences of the denoisers:
\begin{subequations} \label{eq:alpha}
\begin{equation}
    \alpha_0^{(t)} = \frac{1}{n} \sum_{i=0}^{n-1} \frac{\partial G_0(\bs{z}_0^{(t)}, \gamma_0^{(t)})}{\partial z_0^{(t)}[i]}, \;
    \alpha_1^{(t)} = \frac{1}{n} \sum_{i=0}^{n-1} \frac{\partial G_1(\bs{z}_1^{(t)}, \gamma_1^{(t)})}{\partial z_1^{(t)}[i]}.
\end{equation}
\end{subequations}
The choice of denoisers depends on the desired estimation criterion. Here, we focus on the minimum mean square error (MMSE) denoisers, where the goal is to approximate the MMSE estimate \( \wh{\bs{x}} = \Exp[\bs{x} \mid \bs{y}] \).

\paragraph*{Spectral denoiser}  
For MMSE estimation, the optimal spectral denoiser \( G_0(\cdot) \) in~\eqref{eq:mlvamp_g0}, is given by: 
\begin{align} 
    \wh{\bs{x}} = G_0(\bs{z}_0, \gamma_0)
    = \mathbb{E}\!\left[ \bs{x} \mid \bs{z}_0 = \bs{x} + \CNorm\!\left(0, \gamma_0^{-1} \bs{I}\right) \right], \label{eq:gin}
\end{align}
where the expectation is w.r.t. the prior on \( \bs{x} \). Thus, \( \bs{z}_0 \) is a noisy observation of \( \bs{x} \), and \( \wh{\bs{x}} \) its conditional mean. In the interference filtering problem, \( \bs{x} = \sum_\ell \bs{x}_\ell \) with Gaussian components as in~\eqref{eq:varx}. Since the intervals \( B_\ell \) are disjoint~\eqref{eq:Bdisjoint}, the conditional mean~\eqref{eq:gin} reduces to:
\begin{equation} \label{eq:gin_lin}
    \wh{x}[i] =
    \begin{cases}
    \mu_\ell[i] + \dfrac{\gamma_0 S_\ell}{1 + \gamma_0 S_\ell}\big(z_0[i]- \mu_\ell[i]\big), & i \in B_\ell, \\
    0, & i \notin B_\ell,
    \end{cases}
\end{equation}
for all $\ell$. The parameter $\alpha_0$ in~\eqref{eq:mlvamp_g0} then has the closed form:
\begin{equation} \label{eq:alpha_lin}
    \alpha_0 
    = \bkt{G_0'(\bs{z}_0, \gamma_0)} 
    =  \sum_{\ell=1}^{L} \frac{|B_\ell|}{N} 
      \frac{ \gamma_0 S_\ell}{1 + \gamma_0 S_\ell}.
\end{equation}
For convenience, let us define:
\begin{equation} \label{eq:mu_S}
    \mu[i] =
    \begin{cases}
    \mu_\ell[i], & i \in B_\ell, \\
    0, & \text{otherwise},
    \end{cases}
    \quad
    S[i] =
    \begin{cases}
    S_\ell, & i \in B_\ell, \\
    0, & \text{otherwise}.
    \end{cases}
\end{equation}
With this notation,~\eqref{eq:gin_lin} can be expressed compactly as:
\begin{equation}
    \wh{\bs{x}}
    = \bs{\mu} + \frac{\gamma_0 \bs{S}}{1 + \gamma_0 \bs{S}} 
      \odot \big(\bs{z}_0 - \bs{\mu}\big),
\end{equation}
where $\odot$ denotes elementwise multiplication and the fraction is taken elementwise.

\paragraph*{Nonlinear denoiser}  
For MMSE estimation, the nonlinear denoiser $G_1(\cdot)$ in~\eqref{eq:mlvamp_g1} is given by:
\begin{align}
    \wh{\bs{r}} 
    &= G_1(\bs{z}_1,\gamma_1) \nonumber \\
    &= \Exp\!\left[\, \bs{r} \,\big|\, \bs{y} = \Phi(\bs{r}, \bs{w}),\; 
       \bs{r} \sim \CNorm(\bs{z}_1, \gamma_1^{-1}\bs{I}) \right],
    \label{eq:gout}
\end{align}
i.e., the conditional mean of \( \bs{r} \) given the observation \( \bs{y}=\Phi(\bs{r},\bs{w}) \) and Gaussian prior \( \CNorm(\bs{z}_1,\gamma_1^{-1}\bs{I}) \). Since $\Phi(\bs{r},\bs{w})$ is memoryless~\eqref{eq:nonlin_scal}, denoising is componentwise:
\begin{equation}
    \wh{r}[i] = g_1(z_1[i],\gamma_1), 
    \quad i=0,\ldots,N-1,
\end{equation}
where the scalar denoiser is:
\begin{equation}
    g_1(z_1,\gamma_1) 
    = \Exp\!\left[\, r \,\big|\, y = \phi(r,w),\;
      r \sim \CNorm(z_1,\gamma_1^{-1}) \right].
    \label{eq:gout_sca}
\end{equation}

\subsection{Challenges with ML-VAMP}
The ML-VAMP algorithm has been analyzed extensively in the large-system limit \(N \to \infty\) with random unitary \(\bs{V}\). In this regime, performance is exactly characterized by a \emph{state evolution}~\cite{fletcher2018inference}, which describes the joint distributions of both frequency-domain variables \((x[i],\wh{x}[i])\) and time-domain variables \((r[i],\wh{r}[i])\). In certain cases, ML-VAMP can even be shown to be Bayes optimal~\cite{manoel2017multi}. Practical implementation, however, faces two challenges:
\begin{itemize}
    \item \emph{Nonlinear denoiser complexity}: While \(G_0(\cdot)\) admits a closed form expression~\eqref{eq:gin_lin}, \(G_1(\cdot)\) in \eqref{eq:gout_sca} does not. Computing it requires multidimensional numerical integration. For soft saturation~\eqref{eq:wsoft}, the variables are \(w=(w_a,w_b)\) and \(r\), all complex, leading to integration over six real dimensions, which is computationally prohibitive.
    \item \emph{Stability}: While~\cite{fletcher2018inference,manoel2017multi} provide guarantees for fixed iteration \(t\) as \(N \to \infty\), VAMP-based algorithms are well known to suffer from instability.
\end{itemize}

\subsection{Learned ML-VAMP Algorithm}
We propose a learned ML-VAMP framework that retains the iterative VAMP structure while replacing the intractable analytical denoisers with compact trainable neural networks, similar to \cite{borgerding2017amp} but adapted to VAMP. 

To rewrite \eqref{eq:mlvamp} in a learnable form, consider the spectral denoiser output.  
In place of \eqref{eq:mlvamp_z1}, we use:
\begin{equation} \label{eq:z1beta}
    \bs{z}_1^{(t)} = \bs{V}\herm
    (\beta_0^{(t)} \wh{\bs{x}}^{(t)}
    - \beta_1^{(t)} \bs{z}_0^{(t)}),
\end{equation}
where $\beta_0^{(t)}$ and $\beta_1^{(t)}$ are obtained as averages, $(\beta_0^{(t)}, \beta_1^{(t)}) = \langle \bs{\rho}_0^{(t)} \rangle$, with $\rho_0^{(t)}[i]$ given by a neural network:
\begin{equation}
    \rho_0^{(t)}[i] = f_0(z_0^{(t)}[i], S[i], \mu[i], t, \theta),
\end{equation}
where $\theta$ are network parameters and dependence on $t$ allows a different network per iteration. Defining $\wh{\bs{r}}^{(t)}=\bs{V}\herm\wh{\bs{x}}^{(t)}$, \eqref{eq:z1beta} becomes:
\begin{equation} \label{eq:z1beta2}
    \bs{z}_1^{(t)} =
    \beta_0^{(t)} \wh{\bs{r}}^{(t)}
    - \beta_1^{(t)} \bs{V}\herm\bs{z}_0^{(t)}.
\end{equation}
The update of $\gamma_1^{(t+1)}$ follows from the Wiener gain:
\begin{equation}
    \gamma_1^{(t+1)} = \gamma_0^{(t)} \cdot 
    \bkt{\frac{\gamma_0^{(t)} \bs{S}}{1+\gamma_0^{(t)} \bs{S}}}^{-1}.
\end{equation}

For the nonlinear denoiser, the update of $\bs{z}_0^{(t+1)}$ is:
\begin{align}
    \bs{z}_0^{(t+1)} &= \bs{V}\bs{v}^{(t)}, \\
    (v^{(t)}[i], \rho^{(t)}_1[i]) &= f_1(z_1^{(t)}[i],\gamma_1^{(t)},t,\theta), \\
    \gamma_0^{(t+1)} &= \langle\bs{\rho_1}^{(t)}\rangle,
\end{align}
where $f_1(\cdot)$ is a neural network that outputs $v^{(t)}[i]$ and $\rho^{(t)}_1[i]$.

After \(T\) iterations, the signal estimate \(\hat{\bs{x}}_0\) is obtained by bandpass filtering in the frequency domain:
\begin{equation}
    \wh{\bs{x}}_0 = \bs{m}_{B_0} \odot \wh{\bs{x}}^{(T-1)},
\end{equation}
where \(\bs{m}_{B_0} \in \{0,1\}^N\) is a binary mask selecting the band \(B_0\).  
This retains components in \(B_0\) while suppressing others. With quantization, the same procedure applies, using \(\bs{y}_q\) in place of \(\bs{y}\). The proposed algorithm is summarized in Algorithm~\ref{alg:learned_mlvamp}.

\begin{figure}[!t]
\vspace{-3mm}
\begin{algorithm}[H]
\caption{Learned ML-VAMP}
\label{alg:learned_mlvamp}
\begin{algorithmic}[1]
\REQUIRE Prior mean and variance \( (\bs{\mu}, \bs{S}) \) from~\eqref{eq:mu_S}, learned functions \( f_0, f_1 \), and number of iterations \( T \)
\STATE \textbf{Initialize:} \( \bs{z}_1^{(0)} \gets \bs{V}\herm\bs{\mu},\; \gamma_1^{(0)} \gets \bkt{\bs{S}}^{-1} \)
\FOR{each iteration \( t = 0, \dots, T{-}1 \)}
    \STATE \textit{// Neural denoising}
    \FOR{each \( i = 0, \dots, N{-}1 \)}
        \STATE \( (v^{(t)}[i],\, \rho_1^{(t)}[i]) = f_1 \left( z_1^{(t)}[i],\, \gamma_1^{(t)},\, y[i],\, t,\, \theta \right) \)
    \ENDFOR
    \STATE \( \bs{z}_0^{(t)} = \bs{V} \bs{v}^{(t)} \)
    \STATE \( \gamma_0^{(t)} = \langle\bs{\rho_1}^{(t)}\rangle \)
    \STATE \textit{// Spectral denoising}
    \STATE \( \wh{\bs{x}}^{(t)} = \bs{\mu} + \left( \frac{ \gamma_0^{(t)} \bs{S} }{ 1 + \gamma_0^{(t)} \bs{S} } \right) \odot ( \bs{z}_0^{(t)} - \bs{\mu} ) \)
    \STATE \( \gamma_1^{(t+1)} = \gamma_0^{(t)} \cdot \langle\frac{\gamma_0^{(t)} \bs{S}}{1+\gamma_0^{(t)} \bs{S}}\rangle^{-1} \)
    \STATE \textit{// Message update}
    \FOR{each \( i = 0, \dots, N{-}1 \)}
        \STATE \( \rho_0^{(t)}[i] = f_0 \left( z_0^{(t)}[i],\, \gamma_0^{(t)},\, S[i],\, \mu[i],\, t,\, \theta \right) \)
    \ENDFOR
    \STATE \( (\beta_0^{(t)}, \beta_1^{(t)}) = \langle \bs{\rho_0}^{(t)}\rangle \)
    \STATE \( \bs{z}_1^{(t+1)} = \beta_0^{(t)} \bs{V}\herm \wh{\bs{x}}^{(t)} - \beta_1^{(t)} \bs{V}\herm \bs{z}_0^{(t)} \)
    
\ENDFOR
\RETURN \(\wh{\bs{x}}_0 \gets \bs{m}_{B_0} \odot \wh{\bs{x}}^{(T-1)}\)
\end{algorithmic}
\end{algorithm}
\vspace{-7mm}
\end{figure}

\subsection{Model Complexity}
The main advantage of the learned ML-VAMP algorithm~\eqref{alg:learned_mlvamp} lies in its reduced model complexity. A general learned approach would take three $N$-dimensional inputs, $\bs{\mu}$, $\bs{S}$, and $\bs{y}$, and output an $N$-dimensional estimate $\wh{\bs{x}}$, which can greatly increase the number of parameters; for instance, a fully connected layer requires $O(N^2)$ parameters.

In contrast, learned ML-VAMP requires only two learned functions, $f_0(\cdot)$ and $f_1(\cdot)$. For each iteration $t$ with parameters $\theta$, $f_0(\cdot)$ maps 4 inputs to 2 outputs, while $f_1(\cdot)$ maps 3 inputs to 2 outputs. Hence, the complexity does not scale with $N$.

\begin{figure*}[t]
    \centering
    \includegraphics[width=\linewidth]{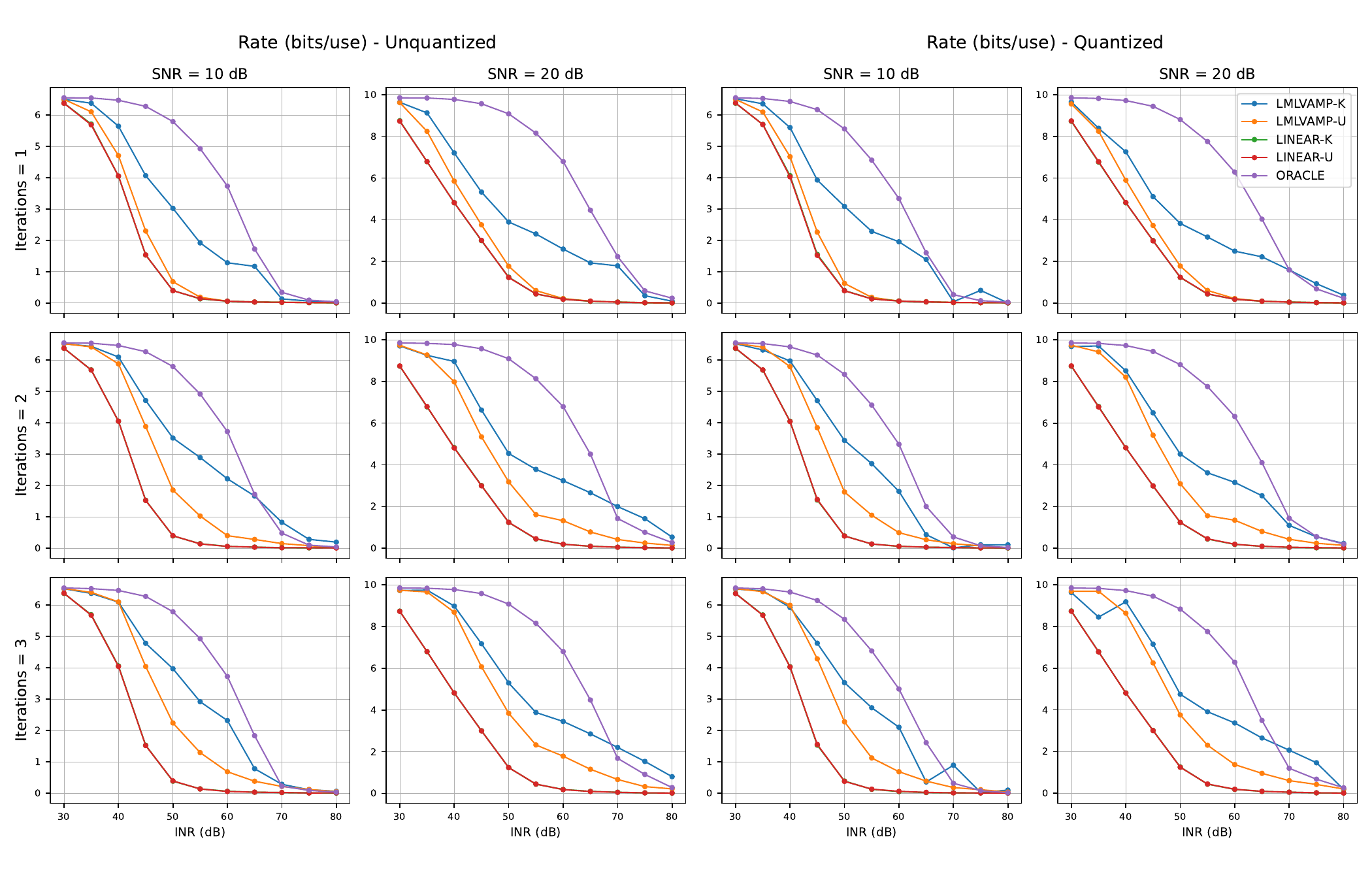}
    \caption{Achievable rate versus INR for different SNRs and VAMP iterations. Left columns represent unquantized, and right columns represent quantized observations. Each row shows a specific number of VAMP iterations at 10~dB and 20~dB SNR.}
    \label{fig:cap_nmse}
\end{figure*}

\subsection{Neural Network Structure}
The learned ML-VAMP framework employs two compact neural networks, $f_0(\cdot)$ and $f_1(\cdot)$, both of which follow a simple two-layer architecture consisting of a hidden fully connected layer with 64 sigmoid units, followed by a linear output layer. Despite their simplicity, these networks effectively capture the required nonlinear dependencies. Both networks operate component-wise, and for notational clarity, the component index $i$ is omitted in the expressions that follow.

\paragraph*{Nonlinear denoiser $f_1(\cdot)$}
At iteration $t$, $f_1$ processes the inputs $(z_1^{(t)}, \gamma_1^{(t)}, y)$, normalized by the saturation level $P_{\text{sat}}$, yielding feature vectors:
\[
\big[ |z_1^{(t)}|/\sqrt{P_{\text{sat}}}, \; (\gamma_1^{(t)} P_{\text{sat}})^{-1}, \; |y|/\sqrt{P_{\text{sat}}} \big].
\]
The outputs are $[v^{(t)}, \rho_1^{(t)}]$, where:
\[
v^{(t)} = z_1^{(t)} + \kappa_0 (y - \kappa_1 z_1),
\quad 
\rho_1^{(t)} = \frac{\gamma_1^{(t)}}{\exp(\log x_{\text{var}})},
\]
with $\kappa_0$, $\kappa_1$, and $\log x_{\text{var}}$ learned from the network. The posterior precision $\gamma_0^{(t)}$ is the componentwise mean of $\rho_1^{(t)}$.

\paragraph*{Message updater $f_0(\cdot)$}
Similarly at each iteration $t$, $f_0$ processes $(z_0^{(t)}, \gamma_0^{(t)}, S, \mu)$, again normalized by $P_{\text{sat}}$, with feature vectors:
\[
\big[ |z_0|/\sqrt{P_{\text{sat}}}, \; (\gamma_0 P_{\text{sat}})^{-1}, \; S/P_{\text{sat}}, \; |\mu|/\sqrt{P_{\text{sat}}} \big].
\]
The network output $\rho_0^{(t)}$ is then averaged across components to yield the message update coefficients $\beta_0^{(t)}$ and $\beta_1^{(t)}$.

\subsection{Training Objective and Loss Design}
The model is trained end-to-end on simulated data, with supervision applied in the frequency domain over the desired band $B_0$. To guide the iterative reconstruction, an intermediate loss penalizes the frequency-domain error at each iteration:
\begin{equation}
\mathcal{L}_{\text{early}} = \sum_{t=1}^{T-1} w_t \left\| \bs{x}[B_0] - \hat{\bs{x}}^{(t)}[B_0] \right\|^2,
\end{equation}
where $w_t = t / \sum_{i=1}^{T-1} i$ are normalized linear weights that assigns greater weight to later iterations. A final loss supervises the last iteration:
\begin{equation}
\mathcal{L}_{\text{final}} = \left\| \bs{x}[B_0] - \hat{\bs{x}}^{(T)}[B_0] \right\|^2.
\end{equation}
The total loss is a weighted combination of both terms:
\begin{equation}
\mathcal{L}_{\text{total}} = \eta \mathcal{L}_{\text{final}} + (1 - \eta) \mathcal{L}_{\text{early}}, \quad \eta \in (0.5,1].
\end{equation}

This loss formulation encourages consistent improvement across iterations while maintaining training stability and convergence. Optimization is performed using the Adam optimizer with an exponentially decaying learning rate.

%% file: 4_performance.tex
\section{Performance Evaluation}
\subsection{Simulation Setup}
Our proposed estimator is evaluated under varying interference levels. The key system-level power ratios, defined relative to the noise variance $\sigma_a^2$, are:
\begin{equation}
\mathrm{SNR} = \frac{S_0 |B_0|}{N \sigma_a^2}, \quad
\mathrm{INR} = \frac{S_1 |B_1|}{N \sigma_a^2}, \quad
\mathrm{SatNR} = \frac{P_{\text{sat}}}{\sigma_a^2},
\end{equation}
corresponding to the signal, interference, and saturation-to-noise ratios. Signal powers are scaled to achieve the desired SNR and INR levels. The quantizer $\mathcal{Q}(\cdot)$ is characterized by its resolution $b$ (in bits) and input backoff $\text{BO}$. Results are averaged over multiple Monte Carlo trials, with simulation parameters summarized in Table~\ref{tab:sim_setup}.

\begin{figure}[!t]
\vspace{-3.6mm}
\begin{table}[H]
\centering
\caption{Simulation Parameters}
\label{tab:sim_setup}
\renewcommand{\arraystretch}{1.2}
\setlength{\tabcolsep}{12pt}
\begin{tabular}{ll ll}
\hline
\textbf{Parameter} & \textbf{Value} & \textbf{Parameter} & \textbf{Value} \\
\hline
$N$ & 512 & $T$ & 1--3 \\
$B_0$ & $[0, 100)$ & $N_{\text{samples}}$ & 1000 \\
$B_1$ & $[300, 400)$ & $N_{\text{epochs}}$ & 2000 \\
$\sigma_a^2$ & 0 dB & $\eta$ & 0.75 \\
$\sigma_b^2$ & --10 dB & \( \mathcal{Q}(\cdot) \) & 10-bit \\
SNR & 10--20 dB & BO & 12 dB \\
INR & 30--80 dB & SatNR & 40 dB \\
\hline
\end{tabular}
\end{table}
\vspace{-5mm}
\end{figure}

\subsection{Evaluated Methods}
To assess the performance of various estimation approaches, we evaluate and compare the following estimators:
\begin{itemize}
    \item \textit{LMLVAMP-K/U:} Learned ML-VAMP, with or without knowledge of the interferer.
    \item \textit{LINEAR-K/U:} Linear frequency-domain Wiener estimator, serving as a lower-bound baseline.
    \item \textit{ORACLE:} Idealized estimator with perfect knowledge of system gain, providing a non-achievable upper bound.
\end{itemize}

\subsection{Evaluation Metrics}
The performance of each estimator is assessed using the following metrics:  
\begin{itemize}
    \item \textit{Achievable Rate Lower Bound:}  
    Computed using the magnitude of the correlation coefficient \( \rho \) between \( \bs{x}_0 \) and \( \hat{\bs{x}}_0 \):
    \begin{equation}
    \rho = \frac{ \left| \mathbb{E} \left[ (\hat{\bs{x}}_0 - \mathbb{E}[\hat{\bs{x}}_0]) (\bs{x}_0 - \mathbb{E}[\bs{x}_0])^* \right] \right| }
    { \sqrt{ \mathrm{Var}(\hat{\bs{x}}_0) \mathrm{Var}(\bs{x}_0) } },
    \end{equation}
    The achievable rate $C$ is lower-bounded as $C \geq -\log_2(1-\rho)$.

    \item \textit{Normalized Mean Squared Error (NMSE):}  
    Measures the relative reconstruction error:
    \begin{equation}
        \mathrm{NMSE} = \mathbb{E} \left[ \frac{ \| \hat{\bs{x}}_0 - \bs{x}_0 \|^2 }{ \| \bs{x}_0 \|^2 } \right].
    \end{equation}
\end{itemize}

\subsection{Quantitative Results}
Fig.~\ref{fig:cap_nmse} summarizes estimator performance. The proposed LMLVAMP-K consistently achieves the highest achievable rate, approaching the oracle benchmark within two iterations, even under high interference. LMLVAMP-U significantly outperforms linear baselines, demonstrating generalization to unknown interference.

Linear estimators degrade sharply at INR levels above 50~dB due to nonlinear spectral leakage, whereas the proposed methods maintain strong performance, with gaps exceeding 20~dB in some cases. Even under 10-bit quantization, LMLVAMP retain a substantial advantage over linear alternatives. The benefit of iterative inference is evident as performance improves noticeably across iterations, particularly for LMLVAMP-U. Furthermore, fixing $\beta_{0}=1$ and $\beta_{1}=0$ across all iterations improves stability while maintaining comparable performance.

%% file: 5_conclusion.tex
\section{Conclusion}
We proposed a Learned ML-VAMP framework for signal recovery in nonlinear RF receivers with strong OOB interference and quantization, impairments commonly encountered in FR3 receiver front-ends. By integrating spectral priors with a trainable denoiser, the method combines model-based interpretability with data-driven adaptability. Its iterative structure alternates between nonlinear denoising and spectral filtering. Simulations show consistent gains over linear baselines in achievable rate, particularly under severe interference. These results highlight the effectiveness of hybrid inference for robust receiver operation in FR3 deployments. Future work includes extensions to time-varying interference, memory effects, and real-time implementations.